\documentclass[twocolumn,preprintnumbers]{revtex4-2}
\usepackage{amssymb}
\usepackage{amsmath}
\usepackage{lineno}
\usepackage{graphicx}
\usepackage{soul}
\usepackage[colorlinks=true,citecolor=blue,linkcolor=blue]{hyperref}
\usepackage[usenames]{color}

\begin{document}

\title{Mechanism of circular polarization in giant pulses and fast radio bursts}
\author{Hui-Chun Wu}
\thanks{huichunwu1@gmail.com, huichunwu@zju.edu.cn}
\affiliation{IFTS, School of Physics, Zhejiang University, Hangzhou 310058, China}

\begin{abstract}
Some giant pulses and fast radio bursts exhibit notable circular polarization, which remains unexplained and carries significant implications for their emission mechanisms. In this study, we identify multiple nanoshot pairs uniformly spaced by approximately 21 $\mu$s within a giant pulse emitted by the Crab pulsar. Among these pairs, a subset displays left-hand and right-hand circular polarization in two distinct nanoshots. We propose that the occurrence of such nanoshot pairs with dual circular polarizations arises from the fragmentation of a linearly-polarized nanoshot along the magnetic field lines under the extreme Faraday effect, leveraging highly-asymmetrical pair plasma and the ultra-intense field of nanoshots. The asymmetry in pair plasmas is likely linked to discharge activities in pulsars. Moreover, the intense field of nanoshots induces cyclotron resonance within the magnetosphere, effectively slowing down the propagation velocity of the circularly polarized mode. Our findings suggest that Crab giant pulses composing nanoshots originate in its polar cap region and escape the magnetosphere along the polar magnetic field. This mechanism can also elucidate the origin of circular polarization in some fast radio bursts and thus lends support to their magnetospheric origin.
\end{abstract}

\keywords{Radio pulsars(1353); Radio bursts(1339); Radio transient sources (2008); Plasma astrophysics (1261)}

\maketitle

\section{Introduction}

Since the discovery of pulsars over half a century ago \citep{Bell}, the physical mechanisms governing their radio emission have remained a profound enigma \citep{Melrose}. This puzzle deepens with the advent of fast radio bursts (FRBs) \citep{Lorimer}, which surpass pulsar radiation in magnitude by several orders. Recently, an FRB event has been linked to a galactic magnetar \citep{FRB-milk-1,FRB-milk-2}. Both pulsars \citep{Philippov} and magnetars \citep{Kaspi} are varieties of rotating neutron stars, with the latter boasting a significantly strong magnetic field. Remarkably, 18 pulsars \citep{Malov} have been observed to sporadically emit giant pulses \citep{giant-cp-1,giant-cp-2,Hankins2003,Soglasnov1,Soglasnov2,Hankins2007,giant-cp-3,Hankins2015,Hankins2016,Eilek}, rivaling the power of FRBs. Some of giant pulses are resolved as clusters of nanosecond bursts, termed ``nanoshots'' \citep{Hankins2007}.

The radiation emitted by pulsars, including their giant pulses, is generally thought to originate within the magnetosphere and propagates along magnetic axes. While the exact source location of FRBs remains contentious, mounting evidence supports a magnetospheric origin \citep{Zhang2023}. Numerous hypotheses have been proposed to explain pulsar radiation, giant pulses, and FRBs, yet none have gained universal acceptance. A pivotal challenge lies in the substantial uncertainty surrounding pair plasmas within the magnetosphere, due to incomplete understanding of the magnetosphere itself \citep{Melrose-unsolved,Kaspi,Philippov}. Establishing a robust magnetosphere model is thus crucial for comprehending these emissions. Radio signals from observations offer limited insights into pair plasmas, underscoring the importance of extracting valuable information from observed signals to refine magnetosphere models and elucidate their radiations.

Our investigation find that a giant pulse emitted by the Crab pulsar contains circularly-polarized (CP) nanoshot pairs with a uniform interval of $\sim21$ $\mu$s. To elucidate this extraordinary phenomenon, we propose an extreme Faraday effect occurring within a highly-asymmetrical pair plasma along the magnetic field lines. This effect directly splits a linearly-polarized nanoshot into a pair of left-circularly polarized (LCP) and right-circularly polarized (RCP) bursts, dictated by their different propagation velocities. Although asymmetrical pair plasmas have long been invoked to explain heightened CP degrees at higher frequencies in pulsar radiation \citep{Hoensbroech}, the notion of highly-asymmetrical pair plasmas has often been met with skepticism. The existence of nanoshot pairs provides compelling evidence for the pronounced asymmetry of pair plasmas, likely stemming from active discharges within the pulsar's magnetosphere. This CP mechanism also applies to FRBs \citep{Dai}, suggesting their potential origin within magnetar magnetospheres.

\section{Nanoshot pairs from the Crab}

The Crab pulsar \citep{Buhler}, initially discovered through its giant pulses \citep{Staelin}, is situated 2 kpc away and rotates with a period of 33.1 ms. Nanoshot pairs were identified within a single giant pulse lasting about 100 $\mu$s (Fig. 5 in Appendix A), as reported by Hankins \textit{et al.} \citep{Hankins2003}. This pulse has a central frequency of 5.5 GHz and a bandwidth of 0.5 GHz. Nanoshot pairs are designated as Pair $(i,i')$, with their respective time ($t$), flux ($S$), and CP sign recorded in Table 1. Pairs $(2,2')$ and $(3,3')$ were initially identified through visual inspection, aided primarily by their distinct CP signs. Motivated by this observation, I measured additional pairs with separations approximating 20 $\mu$s among the high-flux spikes. As a result, the pairs listed in Table 1 demonstrate a relatively uniform spacing, with an average separation of 20.70±0.79 $\mu$s. Notably, the CP signs of the two nanoshots in pairs $(2,2')$ and $(3,3')$ are inverted, transitioning from left hand to right hand. Furthermore, most pairs demonstrate comparable intensities. These features indicate strong correlation between the two nanoshots within each pair.

\begin{table}
\caption{Data of nanoshot pairs extracted from Ref. \citep{Hankins2003},
including time $t$, flux $S$, CP sign (L for
LCP and R for RCP). Pair labels $(i,i')$ are marked on the original
figure (see Fig. 5). The RCP spike $2''$ just behind $2'$ has $t_{2''}=68.71\,\mu\mathrm{s}$ and $S_{2''}=2509.63$Jy.}
\label{tab:1}
\begin{tabular}{p{10pt}|p{26pt}p{26pt}c|p{35pt}p{35pt}|cc}

\hline
$i$ & $t_{i}$($\mu$s) & $t_{i'}$($\mu$s) & $\Delta t_{i,i'}$($\mu$s) & $S_{i}$(Jy) & $S_{i'}$(Jy) & CP$_{i}$  & CP$_{i'}$\\ 
\hline
1 & 18.31 & 37.79 & \bf{19.48} & 552. 59 & 369.84\\
2 & 46.04 & 67.32 & \bf{21.28} & 1660.14 & 2269.14 & \bf{L} & \bf{R}\\
3 & 58.62 & 78.62 & \bf{20.00} & 1023.02 & 1236.40 & \bf{L} & \bf{R}\\
4 & 62.61 & 83.68 & \bf{21.07} & 318.63 & 289.01\\
5 & 75.94 & 96.92 & \bf{20.98} & 1043.11 & 809.65\\
6 & 82.39 & 103.01 & \bf{20.62} & 1287.11 & 728.31\\
7 & 86.19 & 107.74 & \bf{21.55} & 816.68 & 450.67\\
8 & 93.01 & 112.58 & \bf{19.57} & 525.48 & 268.42\\
9 & 11.32 & 33.09 & \bf{21.77} & 166.51 & 928.13 &  & \bf{R}\\ 
\hline
\end{tabular}
\end{table}

An intriguing substructure is also observed within Pair $(2,2')$. Spike $2'$ is immediately followed by $2''$, both exhibiting RCP and separated by $1.39\,\mu$s. Notably, the burst 2 contains two LCP spikes with an ultrashort interval of $\sim10$ ns. Thus, Pair $(2,2'|2'')$  represents a twin pair in close proximity.

The luminosity of nanoshots can be estimated using $L\approx1.2\times10^{30}(\frac{S}{1\mathrm{Jy}})(\frac{R}{1\mathrm{kpc}})^{2}(\frac{\Delta\nu}{\mathrm{1GHz}})$ erg/s \citep{Zhang2023}, where $R=2$ kpc and $\Delta\nu=0.5$ GHz. The electric field is given by $E=\sqrt{L/cr^{2}}$, where $r$ represents the emitter distance and $c$ is the speed of light. Normalizing the field as $a_{0}=eE/m_{e}c\omega_{0}$, with $\omega_{0}=2\pi\nu_{0}$, $e$ as the elementary charge, and $m_{e}$ as the electron mass, yields $a_{0}=45.47(\frac{S}{1\mathrm{Jy}})^{0.5}(\frac{r}{1\mathrm{km}})^{-1}$. Assuming $r=1$ km, $a_{0}=455$ and $2033$ for $S=100$ and $2000$ Jy, respectively. The most intense giant pulse has $S\gtrsim2\times10^{6}$ Jy \citep{Soglasnov2,Hankins2007}, resulting in $a_{0}\approx6.4\times10^{4}$.

It is important to note that a comprehensive statistical analysis of the dataset in \citep{Hankins2003} is constrained by the limited availability and quality of the data. Improved correlation analyses could potentially reveal additional dominant pair intervals (see Appendix A). Future studies with higher-quality data will be necessary to confirm the existence of nanoshot pairs with strong statistical significance.

\section{Extreme Faraday effect in pulsars}

We employ the extreme Faraday effect to elucidate the behavior of nanoshot pairs. This effect was initially proposed to generate CP laser pulses from linearly-polarized lasers within plasmas \citep{Weng}. It relies on the different group velocities of LCP and RCP waves in electron-ion plasmas along the ambient magnetic field. The dispersion relations for LCP and RCP waves are given by $N_{L}^{2}=1-\omega_{pe}^{2}/[\omega_{0}(\omega_{0}+\omega_{B})]$ and $N_{R}^{2}=1-\omega_{pe}^{2}/[\omega_{0}(\omega_{0}-\omega_{B})]$ \citep{Gurnett}, respectively, where $N=k_{0}c/\omega_{0}$ represents the refractive index at wavenumber $k_{0}$ and frequency $\omega_{0}$, $\omega_{pe}^{2}=4\pi e^{2}n_{e}/m_{e}$ denotes the electron plasma frequency at density $n_{e}$, and $\omega_{B}=eB/m_{e}c$ signifies the electron cyclotron frequency in the magnetic field $B$. These dispersion relations apply to both whistler waves and free waves in low-temperature plasmas and are presumed to remain invariant as the waves propagate along the magnetic field lines. They lead to group velocities as
\begin{equation}
\frac{v_{g,L}}{c}=\left[1-\frac{\omega_{pe}^{2}}{\omega_{0}(\omega_{0}+\omega_{B})}\right]\left[1-\frac{\omega_{pe}^{2}\omega_{B}}{2\omega_{0}(\omega_{0}+\omega_{B})^{2}}\right]^{-1},
\end{equation}
\begin{equation}
\frac{v_{g,R}}{c}=\left[1-\frac{\omega_{pe}^{2}}{\omega_{0}(\omega_{0}-\omega_{B})}\right]\left[1+\frac{\omega_{pe}^{2}\omega_{B}}{2\omega_{0}(\omega_{0}-\omega_{B})^{2}}\right]^{-1}.
\end{equation}
In pulsars, there are $\omega_{B}\gg\omega_{0}$ and $\omega_{pe}>\omega_{0}$ typically.

It's notable that conventional pair plasmas do not exhibit the Faraday effect, as LCP and RCP waves follow the same dispersion relation $N_{L,R}^{2}=1-\omega_{pe}^{2}/[\omega_{0}(\omega_{0}+\omega_{B})]-\omega_{pe}^{2}/[\omega_{0}(\omega_{0}-\omega_{B})]$. In essence, the pair plasma remains symmetrical for LCP and RCP modes, preventing their decoupling due to identical velocities.

Pair plasmas in pulsars are characterized by a net Goldreich-Julian charge density $\rho_{GJ}\approx-\mathbf{\mathbf{\Omega}}\bullet\mathbf{B\mathrm{/2\pi c}}$ \citep{Goldreich}, where $\mathbf{\Omega}$ is the angular velocity of the rotating pulsar and $\mathbf{B}$ represents the local magnetic field in the magnetosphere. This net charge density causes a net charge number density $n_{GJ}=\rho_{GJ}/e$. Pair-cascade simulations \citep{Timokhin} indicate electron and positron densities as $n_{e,p}\approx Mn_{GJ}$, where the multiplicity factor $M$ ranges from $10^{3}$ to $10^{5}$. Being much smaller than the background density, the net density insignificantly affects the dispersion relation \citep{Gedalin,Wang}.

Regarding the aforementioned nanoshot pairs, we propose that pair plasmas could exhibit high asymmetry in localized regions, facilitating the extreme Faraday effect. This asymmetry may arise from large net charge density or net current. When radio waves co-propagate with streaming plasma, the Faraday effect diminishes compared to rest or counter-streaming plasma (Appendix B). Consequently, asymmetry is heightened when electrons and positrons possess disparate streaming energies.

Highly-asymmetric pair plasmas have been observed in particle-in-cell (PIC) simulations. Discharge near the pulsar's polar cap results in varied spatial distributions of electrons and positrons with rare positrons in electron-rich regions \citep{Philippov-PIC}. Discharging polar gaps lead to different electron and positron momentum distributions \citep{Timokhin2,Timokhin}.  In magnetar's twisted magnetosphere, an electric gap may possibly form around the equator and induce net currents in pair plasmas \citep{Chen}, such as relativistically streaming positrons against rest electrons, or vice versa. These pair-plasma asymmetries are directly linked to pair discharge processes that sustain pair plasmas within the magnetosphere. Asymmetry may also stem from magnetic reconnections around the equator \citep{Kaspi,Philippov}, involving strong currents.

\begin{figure}[t]
\centering
\includegraphics[width=0.45\textwidth]{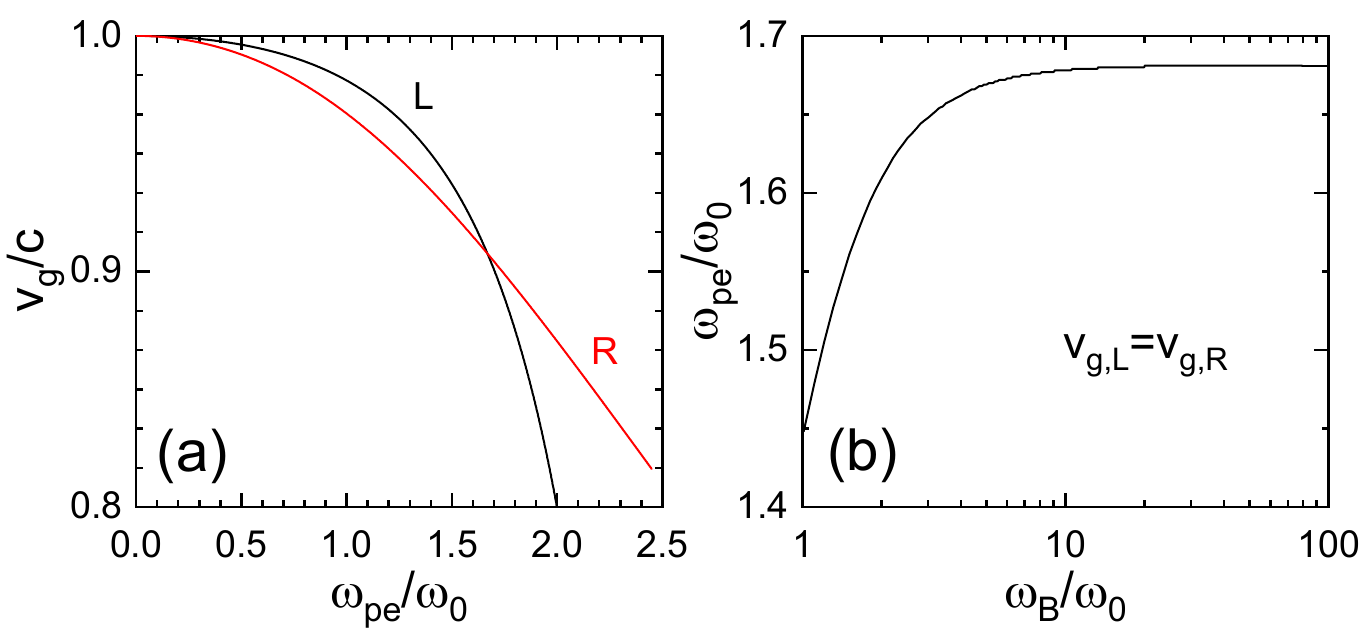} \label{fig1} \caption{(a) Group velocities of LCP and RCP waves in the electron plasmas along the magnetic field at $\omega_{B}/\omega_{0}=5$. (b) The
critical value $\omega_{pe}/\omega_{0}$ for $v_{g,L}=v_{g,R}$.}
\end{figure}

For simplicity, we primarily address the extreme Faraday effect in rest electron plasmas, disregarding positrons due to lower density or higher relativistic factors co-moving with radio waves. We first discuss general features of the extreme Faraday effect in strongly-magnetized regimes $\omega_{B}>\omega_{0}$. Figure 1(a) depicts group velocities of CP waves at $\omega_{B}/\omega_{0}=5$ using Eqs. (1,2). When $\omega_{pe}/\omega_{0}<1.67$, the LCP wave surpasses the RCP wave, explaining nanoshot pairs $(2,2')$ and $(3,3')$ in Table 1. At higher densities with $\omega_{pe}/\omega_{0}>1.67$, the LCP wave lags behind the RCP wave, potentially occurring near the pulsar surface. Figure 1(b) shows that the critical value $\omega_{pe}/\omega_{0}$ for $v_{g,L}=v_{g,R}$ increases with the magnetic field and saturates at $1.68$ for $\omega_{B}/\omega_{0}\gtrsim15$. Results presented in Fig. 1 have been verified by PIC simulations \citep{Deng}.

Figure 2(a) displays group velocities in uniform plasmas with $\omega_{pe}/\omega_{0}=1$, revealing a greater reduction in speed for the RCP mode. The distance required to induce a delay $\Delta t=21\,\mu\mathrm{s}$ between LCP and RCP waves is calculated as $d_{21\mu\mathrm{s}}=c\Delta t/(1-v_{g,R}/v_{g,L})$, equating to 63, 630, and 1260 km for $v_{g,R}/v_{g,L}=0.9$, 0.99, and 0.995, respectively. Given that corotating pair plasmas within the spinning pulsar are contained within the light cylinder radius $r_{lc}=cP/2\pi$, with $P$ as the rotation period, the L-R mode separation must occur within $r\lesssim r_{lc}\approx1579$ km for the Crab. Consequently, the minimum velocity difference should reach $(v_{g,L}-v_{g,R})/v_{g,L}=1-v_{g,R}/v_{g,L}\approx0.005$. Both dimensionless parameters $\omega_{pe}/\omega_{0}$ and $\omega_{B}/\omega_{0}$ govern the ratio $v_{g,R}/v_{g,L}$. Figure 2(b) illustrates the curves for $\omega_{pe}$ and $\omega_{B}$ corresponding to $v_{g,R}/v_{g,L}=$0.9, 0.99 and 0.995. These results indicate that higher magnetic fields reduce the L-R velocity difference for a fixed density.

\begin{figure}[t]
\centering
\includegraphics[width=0.45\textwidth]{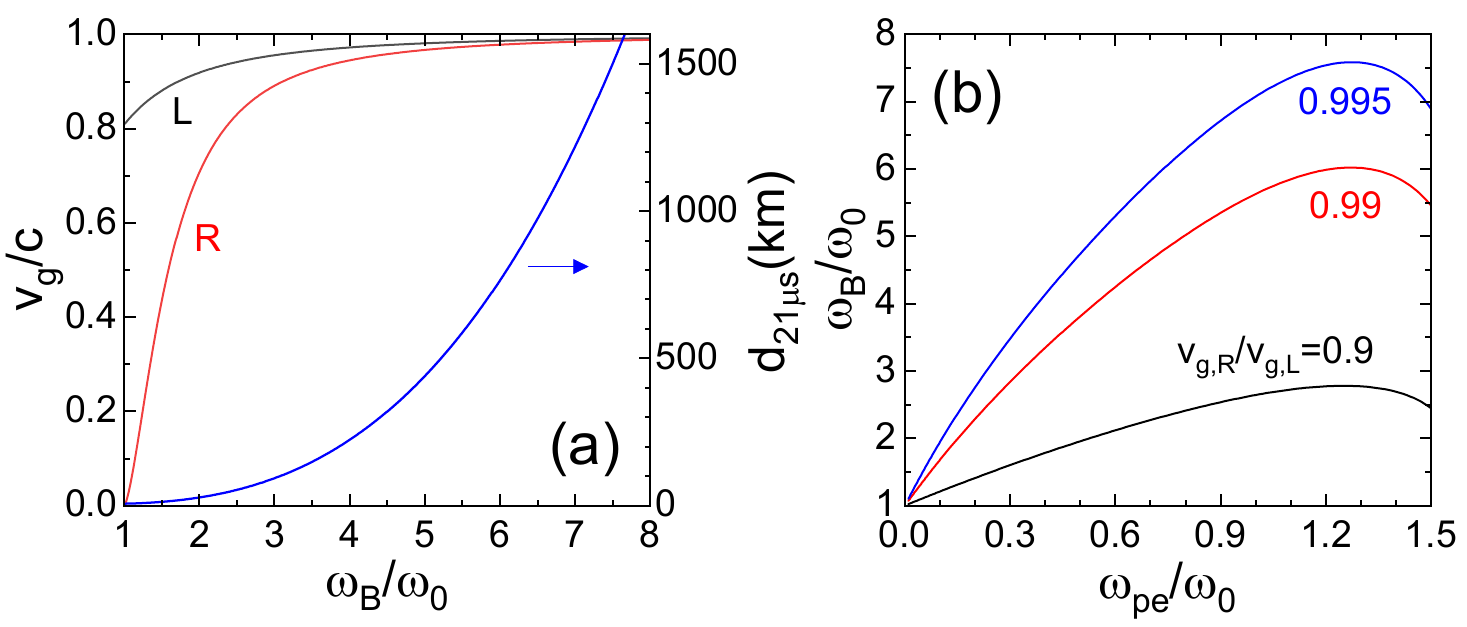} \label{fig2} \caption{(a) Group velocities for LCP and RCP waves at $\omega_{pe}/\omega_{0}=1$. The transmission distance required to produce a $21\,\mu\mathrm{s}$ L-R delay is determined by the ratio $v_{g,R}/v_{g,L}$. (b) Parameter curves for $v_{g,R}/v_{g,L}=0.9, 0.99, 0.995$.}
\end{figure}

\section{Application to nanoshot pairs}

We begin by estimating the delay between left-hand and right-hand CP mode by $\Delta t=c^{-1}\intop_{r_{0}}^{r_{lc}}(1-v_{g,R}/v_{g,L})dr$ induced across the entire Crab magnetosphere for nanoshots at 5.5 GHz. The electron density is $n_{e}=Mn_{DJ}$, with $M=10^{3}$ and $n_{GJ}\approx6.9\times10^{10}\mathrm{cm^{-3}}(\frac{B}{10^{12}\mathrm{Gs}})(\frac{P}{1\mathrm{s}})^{-1}$. The magnetic field follows $B=B_{s}(r/r_{0})^{-3}$, where the Crab's surface magnetic field $B_{s}\approx4\times10^{12}$ Gs and its radius $r_{0}=10$ km \citep{Eilek}. This calculation yields a delay of $\Delta t\approx6\times10^{-5}\mathrm{\,ns}$, which increases to $4\times10^{-3}\mathrm{\,ns}$ with $M=10^{5}$. Consequently, the L-R velocity difference is insufficient to split a linearly-polarized nanoshot into CP modes as described by linear theory.

Considering the ultra-intense field of nanoshots with $a_{0}\gg1$, relativistic effects \citep{Mourou,YangYP} become relevant. The relativistic dispersion relations for strong CP waves in electron-ion plasmas are modified as \citep{Sheng,YangXH,CaoLH}
\begin{equation}
N_{L}^{2}=1-\frac{\omega_{pe}^{2}/\gamma_{e}}{\omega_{0}(\omega_{0}+\omega_{B}/\gamma_{e})},
\end{equation}
\begin{equation}
N_{R}^{2}=1-\frac{\omega_{pe}^{2}/\gamma_{e}}{\omega_{0}(\omega_{0}-\omega_{B}/\gamma_{e})},
\end{equation}
where $\omega_{pe}^{2}$ and $\omega_{B}$ are divided by the electron relativistic factor $\gamma_{e}=\sqrt{1+a_{0}^{2}}\approx a_{0}$. These effects stem from mass corrections of electrons ($m_{e}\rightarrow\gamma_{e} m_{e}$) in plasma and cyclotron frequencies. The effect of plasma frequency reduced by relativistic electron motion in strong fields is called as relativistic-induced transparency.

Additionally, relativistic radiation pressure can efficiently compress pair plasmas \citep{Weng-ppcf}, owing to their much lower mass density compared to electron-ion plasmas. One-dimensional PIC simulations demonstrate that this compression terminates until plasmas become dense enough to reflect incident waves. However, in real space, this relativistic-induced opacity is mitigated by transverse diffusion of pair plasmas. The plasma frequency term can be expressed as 
\begin{equation}
\frac{\alpha\omega_{pe}^{2}}{\gamma_{e}}=\frac{4\pi e^{2}}{m_{e}}\frac{\alpha n_{e}}{\gamma_{e}}=\frac{4\pi e^{2}}{m_{e}}\frac{\alpha Mn_{GJ}}{\gamma_{e}},
\end{equation}
where $\alpha$ is the compression ratio rising with $a_{0}$. We crudely assume $\alpha/\gamma_{e}\approx\alpha/a_{0}\sim O(1)$, i.e. $\alpha\omega_{pe}^{2}/\gamma_{e}\approx\omega_{pe}^{2}$, implying relativistic-induced transparency and opacity balance out in pair plasmas. The factor $M$ varies in
orders of magnitude, and here we take $M=10^{3}$. Although strong nanoshots might induce pair cascades \citep{Zhang}, an aspect not addressed here.

Subsequently, we calculate group velocities of nanoshots by substituting $\omega_{B}\rightarrow\omega_{B}/a_{0}$ in Eqs. (1,2) with Crab parameters. Figure 3 displays group velocities of CP waves at $a_{0}=10000$, 1000, 450, and 100. We maintain a constant $a_{0}$ due to several reasons. Firstly, the localized emitter and highly-asymmetric pair plasmas near the emitter within the light cylinder suggest local interpretations of results in Fig. 3, rather than a simple diffractionless wave propagation from the pulsar surface to the light-cylinder boundary. Secondly, relativistic emissions may highly beam nanoshots, and the relativistic self-focusing effect \citep{Mourou} could counteract diffraction.

\begin{figure}[t]
\centering
\includegraphics[width=0.48\textwidth]{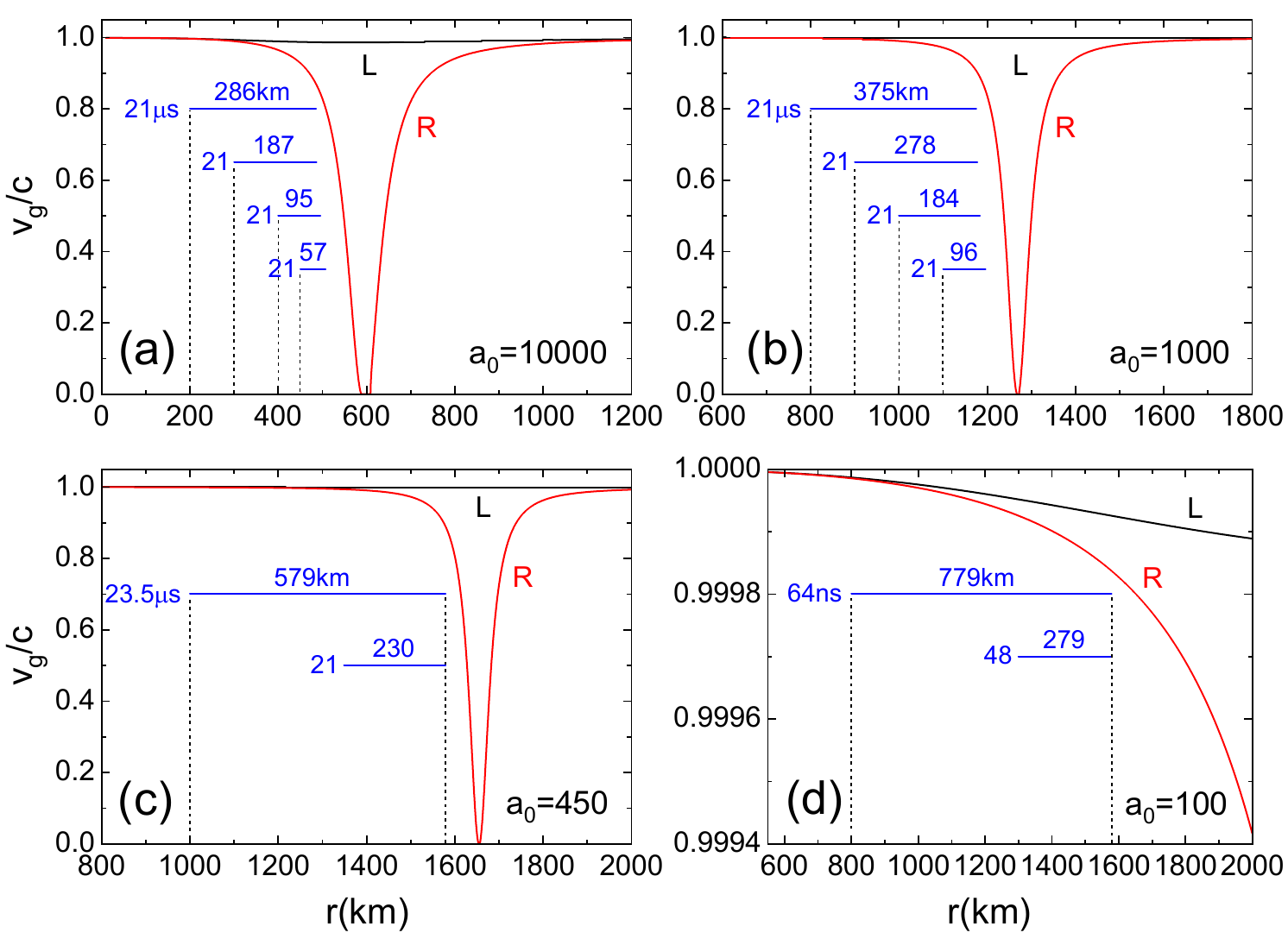} \label{fig3} \caption{Group velocities of LCP and RCP waves recorrected by relativistic effects of nanoshots. (a) $a_{0}=10000$. Blue lines are selected paths (starting from $r_{0}=200$, 300, 400 and 450 km) inducing the $21\,\mu\mathrm{s}$ delay for LCP and RCP waves. Vertical dashed lines mark the starting points. The path length is shown also. (b) $a_{0}=1000.$ The $21\,\mu\mathrm{s}$-delayed paths start from $r_{0}=800$, 900, 1000 and 1100 km. (c) $a_{0}=450$. The $21\,\mu\mathrm{s}$-delayed path starts from $r_{0}=1349$ km and ends at the light cylinder $r_{lc}\approx1579$ km. The second path is from $r_{0}=$1000 km to $r_{lc}$. (d) $a_{0}=100.$ Both nanosecond-delayed paths end at $r_{lc}$.}
\end{figure}

In Fig. 3, the LCP wave consistently propagates near the light speed $c$. The velocity dip of the RCP wave indicates resonance with electron cyclotron motion at $\omega_{B}/a_{0}\approx\omega{}_{0}$. This resonance point shifts outward for weaker fields, extending beyond the light cylinder at $a_{0}=450$. The $21\,\mu\mathrm{s}$-delayed path can be as short as tens of kilometers near the resonance point due to significantly reduced RCP velocity there. The $21\,\mu\mathrm{s}$ delay remains within the light cylinder for $a_{0}=450$. At $a_{0}=100$, the delay reduces to tens of nanoseconds but still exceeds nanoshot widths. The superposition of sub-$\mu\mathrm{s}$-separated nanoshots explains the partial CP state of $\mu$s-scale giant pulses \citep{giant-cp-1,giant-cp-2,giant-cp-3,Hankins2016}, typically exhibiting opposite CP signs at the front/tail and remaining linearly polarized at the center. Since RCP velocity is sensitive to field strength, uniform delay suggests original linearly-polarized nanoshots share similar intensities.

Pair $(2,2'|2'')$ is identified as a twin pair, with spikes $2'$ and $2''$ separated by $1.39\,\mu\mathrm{s}$. A fluctuation in RCP velocity due to 0.5 GHz bandwidth (9\% relatively) can explain this interval. Assuming spike $2''$  has a slightly higher central frequency at $1.03\omega_{0}$, Figure 4 depicts group velocities and final delays of two RCP waves at $\omega_{0}$ and $1.03\omega_{0}$ at the end of $21\,\mu\mathrm{s}$-delayed paths from Fig. 3. Notably, RCP waves performs a negative dispersion and their delay correlates positively with $v_{g,R(\omega_{0})}/v_{g,R(1.03\omega_{0})}$ at the endpoint of paths. The results of $a_{0}=10000$ reproduce the observed value.

The slowing region of the RCP wave expands with increasing particle multiplicities. For a multiplicity factor of $M=10^4$, the resonance pit of $v_{g,R}$ in Fig. 3(c) widens by approximately threefold, and the propagation from $r_0 = 1349$ km to $r_{lc}$ results in a delay of 142 $\mu\mathrm{s}$, significantly exceeding the previous 21 $\mu\mathrm{s}$. Likewise, for $M=10^4$, the L-R delays along the same paths in Fig. 3(d) increase by an order of magnitude, reaching hundreds of nanoseconds. Consequently, a higher plasma density facilitates Faraday splitting at comparatively weaker giant-pulse fields.

\begin{figure}[t]
\centering
\includegraphics[width=0.45\textwidth]{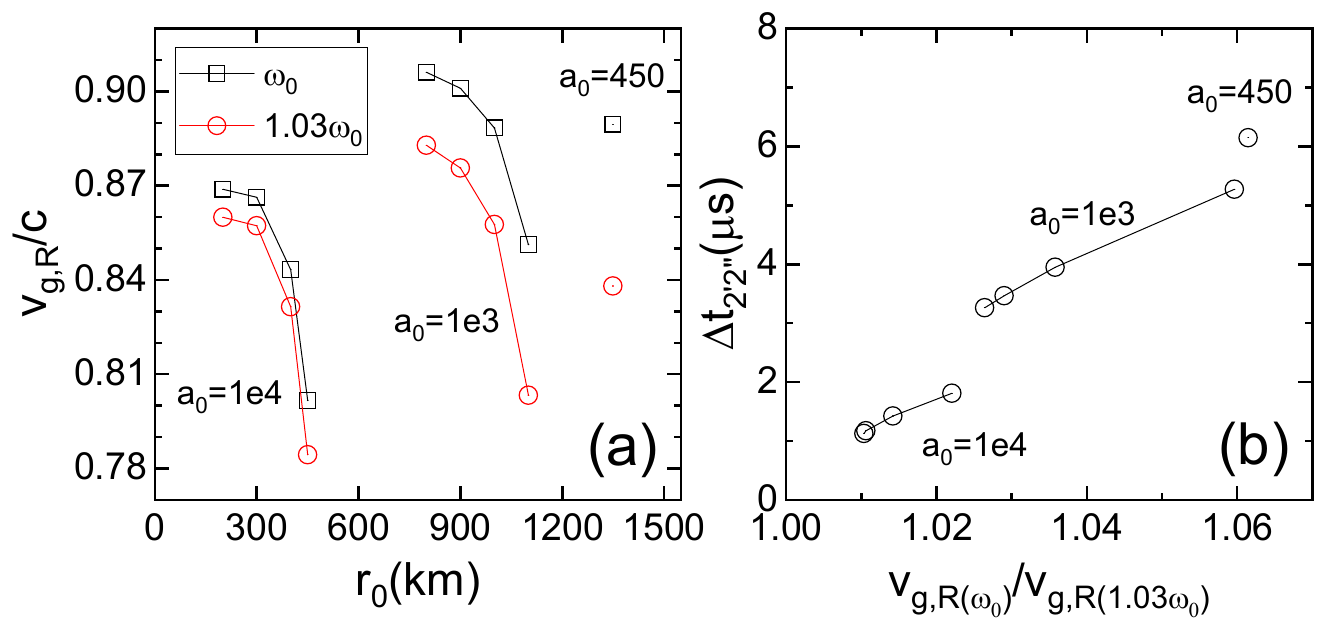} \label{fig4} \caption{(a) Group velocities of RCP waves with the frequencies $\omega_{0}$ and $1.03\omega_{0}$ at the end of all the $21\,\mu\mathrm{s}$-delayed paths in Fig. 3. They are plotted as a function of the path starting position $r_{0}$. (b) The time delay of RCP waves at $\omega_{0}$ and $1.03\omega_{0}$ as a function of ratio of their velocities at the path end given in (a).}
\end{figure}

We do not consider L-R separation during or after the resonance point, as asymmetric plasmas might be localized, thus limiting this process. At resonance, symmetrical plasmas likely dampen both LCP and RCP nanoshots significantly, explaining comparable intensities in most nanoshot pairs. After experiencing energy loss of resonance, nanoshot pairs will propagate with a weaker field, as shown in Fig. 3(d), without additional separation. Further research are required on the interaction between strong waves and pair plasmas within or outside the light cylinder to substantiate intact propagation of nanoshot pairs.

It's worth noting that another Crab giant pulse at $43\pm1$ GHz exhibits negative dispersion \citep{Hankins2016}, where the 42 GHz component leads the 44 GHz one by $\sim2$ $\mu\mathrm{s}$. This delay is attributable to slower CP mode at higher frequency before resonance in the magnetosphere (wave dispersion is positive after the resonance point). Importantly, interstellar propagation cannot produce such negative dispersion.

Lastly, we discuss Faraday rotation before complete L-R mode decoupling. Nanoshot centers remain linearly-polarized, with polarization angle rotation with distance as \citep{Gurnett}
\begin{equation}
\theta=\frac{1}{2}\intop(k_{0,L}-k_{0,R})dr=\frac{\omega_{0}}{2c}\intop(N_{L}-N_{R})dr.
\end{equation}
While Faraday rotation is typically considered in the interstellar region ($N_{L,R}<1$), our scenario with $N_{L}<1$ and $N_{R}>1$ before resonance implies faster rotation, sensitive to local plasma density. This accounts for observed random polarization angles in some nanoshots \citep{Hankins2016}.

\section{Discussions and conclusions}

Aside from the nanoshot pairs shown in Fig. 5, we did not find other giant pulses comprising nanoshot pairs in the literature. Therefore, this study highlights the need for further observations of giant pulses at $\sim1$ ns resolution for the Crab and other pulsars, or targeted searches in relevant databases. Nanoshot pairs should be readily identifiable in giant pulses with sparse and distinct nanoshots, particularly with the aid of opposite CP signs. The event discussed here was observed in the main-pulse component \citep{Hankins2003} of the Crab's mean profile \citep{Hankins2015}. For simplicity, we focus on the scenario where the intersection angle between the wave vector and the magnetic field is exactly zero. For pair plasmas streaming along the magnetic field, a small intersection angle can become significant when Lorentz transformed into the rest frame of the plasmas \citep{Hoensbroech}, thereby limiting the Faraday effect. This stringent requirement suggests that nanoshot pairs might rarely be seen in the main-pulse component of the Crab. Since reconnection sites involve twisted magnetic fields in the equatorial region \citep{Parfrey,Philippov1}, they are unlikely to generate nanoshot pairs. Therefore, the Crab's main-pulse component, including its giant pulses, is more likely produced above its polar cap and propagates along the polar magnetic field.

The Crab pulsar emits seven discrete radio components from various locations within its spin cycle \citep{Hankins2015}. Giant pulses are also detected in the interpulse component \citep{Hankins2016}, which is situated approximately $140^\circ$ from the main-pulse component (with a full cycle spanning $360^\circ$). In contrast to the main-pulse case, giant pulses from the interpulse component exhibit no distinct substructures, suggesting that they do not escape the magnetosphere along the polar magnetic field. Consequently, nanoshot pairs are not anticipated within the interpulse component.

\begin{figure*}[t]
\centering
\includegraphics[width=0.8\textwidth]{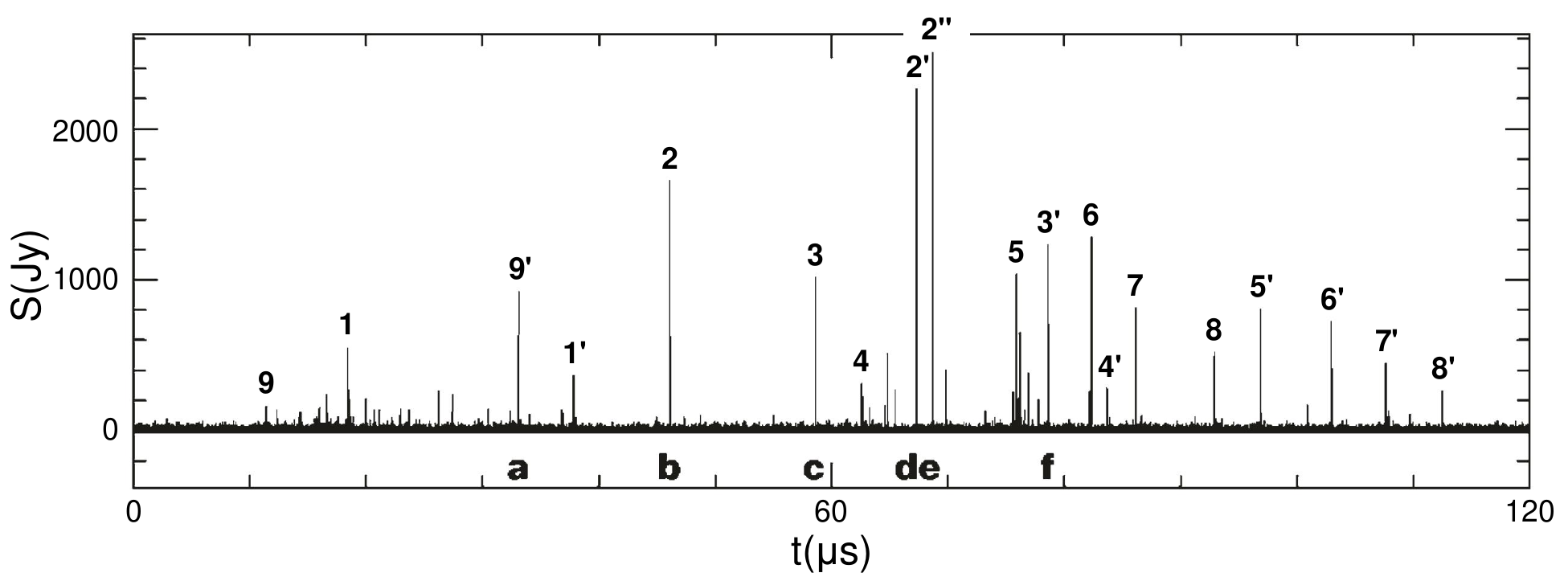} \label{fig5} \caption{Labeled nanoshot pairs $(i,i')$ on the original Fig. 2 in Ref. \citep{Hankins2003}.}
\end{figure*}

Most FRBs exhibit linear polarization \citep{Qu-frb-cp}, with some showing a significant CP degree \citep{FRB-cp-Masui,FRB-cp-Petroff,FRB-cp-Cho,FRB-cp-Day,FRB-cp-Xu,FRB-cp-Kumar}. The much higher magnetic fields in magnetars, compared to pulsars, tend to discourage the CP mode splitting, which may explain the prevalence of linearly polarized FRBs. Events with notable CP can be analogously explained as giant pulses. In Ref. \citep{FRB-cp-Cho}, an FRB displays both LCP and RCP components at its front and tail with linear polarization at center. In Refs. \citep{FRB-cp-Masui,FRB-cp-Petroff,FRB-cp-Day,FRB-cp-Xu,FRB-cp-Kumar}, only one CP state precedes linear polarization in FRBs, possibly due to severe delay and damping of the opposite-sign CP component. Similar to giant pulses composing nanoshots, CP mode splitting in millisecond FRBs likely occurs on much shorter timescales. Indeed, some FRBs have been resolved to contain subpulses lasting from 60 ns to 5 $\mu$s \citep{FRB-nano-Nimmo-1,FRB-nano-Nimmo-2}. Hankins \textit{et al.} \citep{Hankins2003,Hankins2007} suggested that nanoshots could be the fundamental units in giant pulses. By detecting events with nanosecond resolution, one could identify such radiation elements in FRBs, potentially as short as the inverse of the bandwidth. The variable and high Faraday rotation measures observed in a few repeating FRBs \citep{FRB-Michilli,FRB-reverse-Anna} can be explained by extreme Faraday rotation in the magnetosphere as discussed above.

If similar nanoshot-pair phenomena were observed in FRBs, the following arguments would apply. Firstly, dominant plasma-dispersion stretching and transverse magnetic fields would prohibit the extreme Faraday effect in the distant FRB-shock model. Secondly, the Lorentz-transformed enlargement of the intersection angle mentioned above would also hinder CP splitting in the FRB-reconnection model. Therefore, FRBs exhibiting nanoshot pairs would disfavor the shock and reconnection models and should be more consistent with an emission site located in the polar cap region. Similar to the Crab's giant pulses, FRBs can escape freely from the inner magnetosphere along the polar magnetic field. It has been argued that FRBs could be significantly damped by particle scattering in magnetic fields \citep{Beloborodov1,Beloborodov2}. However, a more general analysis does not indicate such severe scattering \citep{Qu}. Specifically, the phenomenon of particle surfing on the burst front, driven by ponderomotive forces, can effectively mitigate substantial damping by particles \citep{Lyutikov}.

By analyzing CP data, one can infer global properties of pair plasmas and magnetic fields. However, there are differing conventions regarding CP sign definitions and measures of CP degree such as the Stokes parameter $V$. We adopt the convention in Ref. \citep{Straten}, where CP handedness is defined from the point of view of the source, with $V>0$ indicating LCP and $V<0$ for RCP. If the CP sign in Ref. \citep{Hankins2003} differs from our convention, then a plasma with a surplus of positrons or a reversed magnetic field is necessary to explain the data. Recently, Anna-Thomas \textit{et al.} \citep{FRB-reverse-Anna} reported that the Faraday rotation of a repeating FRB reversed twice within months, possibly due to changes in charge sign or magnetic field direction in the magnetosphere.

For simplicity, the dispersion relation and the resulting Faraday splitting in the Crab magnetosphere are analyzed under the assumption that the wave vector is perfectly aligned with the magnetic field in a single-species plasma at rest. Deviations from this scenario, such as a finite angle between the wave vector and the magnetic field or the presence of partially-asymmetric double-species plasmas, are anticipated to diminish and constrain Faraday splitting. Additionally, plasma streaming along the magnetic field plays a crucial role. When radio waves propagate precisely along the magnetic field, counter-streaming motion of the plasma (relative to the wave vector) may enhance Faraday splitting (Appendix B), leading to decoupling of CP modes over a shorter distance than previously indicated. Conversely, Faraday effect in co-streaming plasma is less effective compared to plasma at rest. Consequently, nanoshot pairs may offer insights into the streaming motion within the magnetosphere.

In summary, we have identified nanoshot pairs within a giant pulse from the Crab pulsar, attributed to the extreme Faraday effect in highly-asymmetric pair plasmas. This study suggests that giant pulses observed in the Crab's main-pulse component originate in the polar cap region and escape the magnetosphere along the polar magnetic field lines. The proposed mechanism is also applicable to explaining circular polarization and extreme Faraday rotation observed in FRBs. This research calls for further observational confirmation of nanoshot pairs in both giant pulses and FRBs. If validated, as a distinctive phenomenon, nanoshot pairs can serve as a robust diagnostic tool for probing magnetospheric characteristics, which is essential for accurate modeling of magnetospheres and their coherent radio emissions.

\textbf{Appendix}

\textbf{A. Nanoshot-pair labeling}

In Fig. 5, nine pairs of nanoshots are labelled as $(i,i')$ on the original Figure 2 in Ref. \citep{Hankins2003}. The spikes between 4 and $3'$ might represent side pulses of strong nanoshots $2',\,2''$ and 5, thus they do not be paired. The CP states of nanoshots with the tags (a-f) are shown in original Panels (a-f) of the figure. The CP sign of the most intense spike within nanoshots (a-f) is listed in Table 1. The substructure of burst 2 can be observed in Panel b. Pair $(9,\,9')$ is uncertain since spike 9 is much weaker than $9'$. The data were extracted using Getdata Graph Digitizer.

Since the above nanoshot pairings are determined manually, it is essential to perform an objective statistical analysis of the intervals between any two nanoshots. From the original data, we extracted a total of 55 nanoshots with fluxes exceeding 78 Jy. The separation $\Delta t_{i,j}$ between any two spikes $(i,j)$ follows the distribution shown in Fig. 6, with a bin size of $1\,\mu\mathrm{s}$. Each spike is assumed to contribute to only one pair per bin. Notably, $\Delta t_{i,j}=21\,\mu\mathrm{s}$ corresponds to a distinct peak, while a separation of $9\,\mu\mathrm{s}$ also appears dominant. However, the validity of these pairings remains uncertain in the absence of CP data.

\textbf{B. CP mode splitting in streaming plasmas}

It is commonly believed that pair plasmas stream out of pulsars along the magnetic field lines with a relativistic factor $\gamma=100-1000$. Therefore, it's important to discuss the extreme Faraday effect in streaming plasmas. Here we only discuss the case that radio waves propagate exactly along the magnetic field, i.e. plasma-streaming direction. Assuming that CP waves propagate along the $z$ axis in a uniform rest plasma with a length $l$, the delay of L-R waves after crossing this plasma block is $\Delta t=(1-v_{g,R}/v_{g,L})l/c\approx(v_{g,L}-v_{g,R})l/c^{2}\propto v_{g,L}-v_{g,R}$, determined by the velocity difference $\Delta v_{g}\equiv v_{g,L}-v_{g,R}$.

Let's consider a new frame moving along the $-z$ axis with a constant velocity $v$. In this moving frame, the plasma streams along the $z$ axis with velocity $v$, and the CP-wave velocity is given by the transformation $v_{g}^{\prime}=(v_{g}+v)(1+vv_{g}/c^{2})$. Thus, one can deduce $\Delta v_{g}^{\prime}=\Delta v_{g}(1-v^{2}/c^{2})(1+vv_{g}/c^{2})^{-2}\approx\Delta v_{g}/4\gamma^{2}\ll\Delta v_{g}$, where $v\approx c$, $v_{g}\approx c$ and $\gamma=(1-v^{2}/c^{2})^{-1/2}$. Consequently, when CP waves propagate in the co-moving plasma, CP mode splitting is not effective.

If the new frame moves along the $z$ axis with velocity $v$, the plasma will stream along the $-z$ axis, opposite to the wave propagation direction. Then, one $v_{g}^{\prime\prime}=(v_{g}-v)(1-vv_{g}/c^{2})$
and $\Delta v_{g}^{\prime\prime}=\Delta v_{g}(1-v^{2}/c^{2})(1-vv_{g}/c^{2})^{-2}=\Delta v_{g}\tilde{\gamma}^{4}/\gamma^{2}$, where $\tilde{\gamma}=(1-vv_{g}/c^{2})^{-1/2}$. If $v_{g}\geq v\approx c$, then $\Delta v_{g}^{\prime\prime}\geq\Delta v_{g}\gamma^{2}\gg\Delta v_{g}$, which is applicable when $\Delta v_{g}$ is very small in the rest plasma and can be significantly enhanced by the counter-streaming plasma. However, when nanoshots collide head-to-head with electrons or positrons, particles with $\gamma\lesssim a_{0}/2$ will be decelerated to rest within strong fields \citep{Wu}. Thus, Faraday enhancement by the counter-streaming plasma might be difficult for giant pulses and FRBs. It's more likely that counter-streaming plasmas would have low energy in strong fields.

In a plasma outflow containing both low-energy electrons and relativistic positrons, the influence of positrons can be neglected, allowing for the discussion of the Faraday effect primarily within the electron plasma. Consequently, this type of pair plasma, characterized by differing streaming energies of electrons and positrons, exhibits significant asymmetry. Physically, the extent of CP mode splitting and Faraday rotation is determined solely by the relative cross distance between the radio waves and the plasma. This net crossing, occurring between co-traveling waves and plasma, is significantly slower compared to wave propagation in stationary or counter-streaming plasma.

\begin{figure}[t]
\centering
\includegraphics[width=0.45\textwidth]{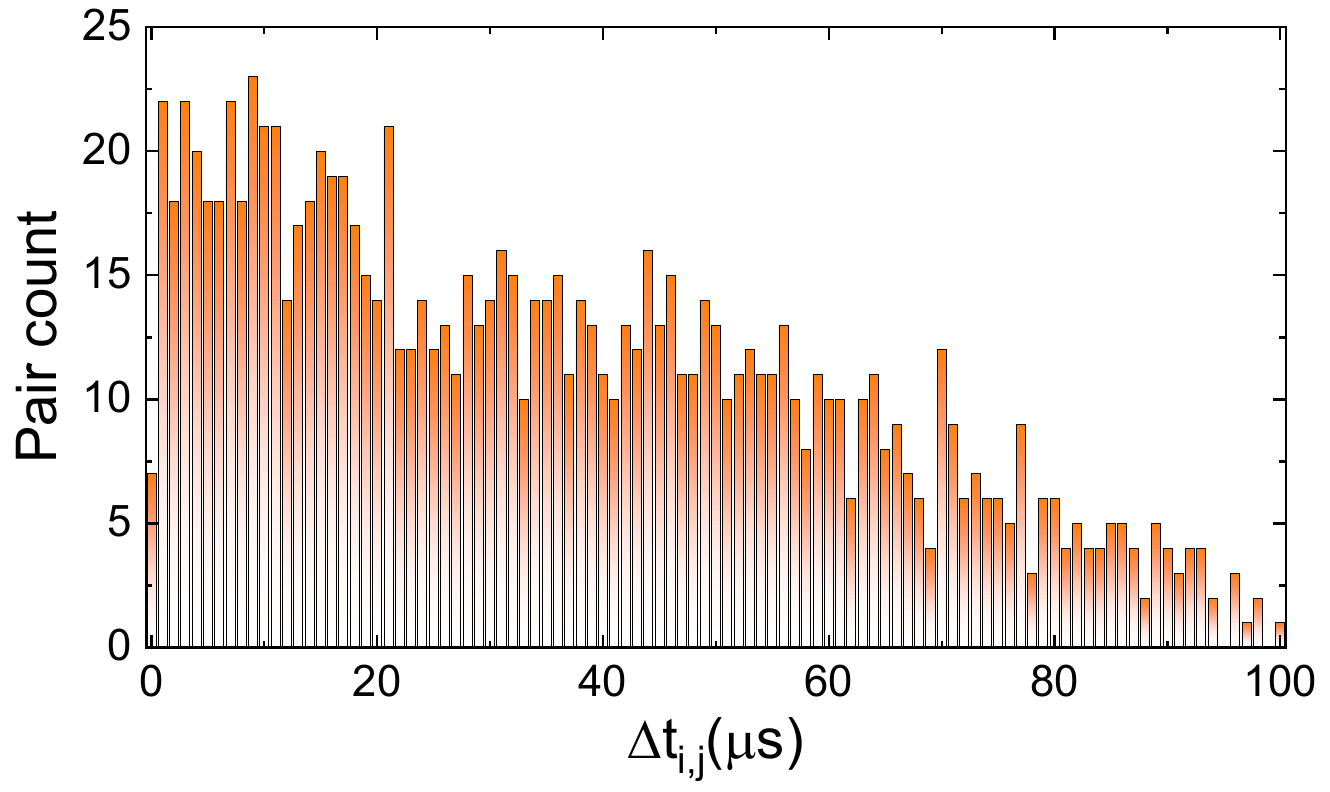} \label{fig6} \caption{Distribution of the spacing of any two spikes above 78 Jy on Fig. 2 in Ref. \citep{Hankins2003}.}
\end{figure}

\medskip

This work was partially supported by the Strategic Priority Research Program of Chinese Academy of Sciences (No. XDA17040502). 

\bibliography{ms}

\end{document}